\documentclass[%
 reprint,
 amsmath,amssymb,
 aps,
]{revtex4-2}

\usepackage{graphicx}
\usepackage{dcolumn}
\usepackage{bm}
\usepackage[dvipsnames]{xcolor} 

\makeatletter
\DeclareRobustCommand\onedot{\futurelet\@let@token\@onedot}
\def\@onedot{\ifx\@let@token.\else.\null\fi\xspace}

\makeatother

\makeatletter
\newcounter{savesection}
\newcounter{apdxsection}
\renewcommand\appendix{\par
  \setcounter{savesection}{\value{section}}%
  \setcounter{section}{\value{apdxsection}}%
  \setcounter{subsection}{0}%
  \gdef\thesection{\@Alph\c@section}}
\newcommand\unappendix{\par
  \setcounter{apdxsection}{\value{section}}%
  \setcounter{section}{\value{savesection}}%
  \setcounter{subsection}{0}%
  \gdef\thesection{\@arabic\c@section}}
\makeatother

\usepackage{hyperref} \hypersetup{colorlinks=true, citecolor = blue, linkcolor=teal, filecolor=magenta, urlcolor=cyan}

\DeclareMathOperator\erf{erf}

\begin{document}

\preprint{APS/123-QED}

\title{Anomalous Shear Stress Growth \\ During Relaxation of a Soft Glass}

\author{Crystal E. Owens}
 \email{crystalo@mit.edu}
\affiliation{Massachusetts Institute of Technology, 77 Massachusetts Ave, Cambridge, USA}


\date{\today}

\begin{abstract}
We show experimentally that multiple soft glassy fluids are capable of storing directional rheological signatures from past shear history, evidenced during \textcolor{black}{stress growth and overall} nonmonotonic stress relaxation after small \textcolor{black}{steps in strain}.
We illustrate theoretically that these responses can be reproduced without requiring thixotropy or shear-banding, which are typically implicated in time-dependent rheological complexities, but by using a simple elastoplastic rheological model with power-law yielding (EP-PLY) that incorporates a distribution of local strain states. 
Using insight from the model, we suggest a mechanism for the experimentally observed stress increase to be driven by residual anisotropy in strain states that are relaxed at different rates. \textcolor{black}{We demonstrate that these effects persist even after material is stressed beyond the yield stress, indicating that past deformation may have more influence than previously thought.}

\end{abstract}

\maketitle

\textcolor{black}{Deformation history imbues soft glassy materials with subtle yet pervasive artifacts embedded within the microstructural state. These can persist even long after previous processing \cite{song2022microscopic, Murphy2020, di2022transient}, and the same material may exhibit varied mechanical responses and mesoscopic phases \cite{PhysRevLett.125.168003, PetekidisTwoStepYielding}. This has wide impact, from food products to pharmaceuticals and soil, where mitigating or at least understanding the origins and evolution of residual stresses is essential to material design, handling, performance, and lifespan. Such influences appear frequently in rheology, where pre-experiment ``pre-shear" at high rates is widely used to set a reproducible initial state. In creep tests, this leads to irregular or even non-monotonic strain responses for applied creep stresses $\sigma$ up to 10-20\% of the yield stress $\sigma_y$ \cite{Coussot2006, Cloitre2000, lidon2017power}. In start-up of flow tests, it leads to stress and strain differences at yielding \cite{Choi2020, edera2025mechanical}. However, the evidence of this influence is typically overlaid upon other protocols, obscuring or changing it.}

\begin{figure}
\includegraphics[width=0.93\columnwidth]{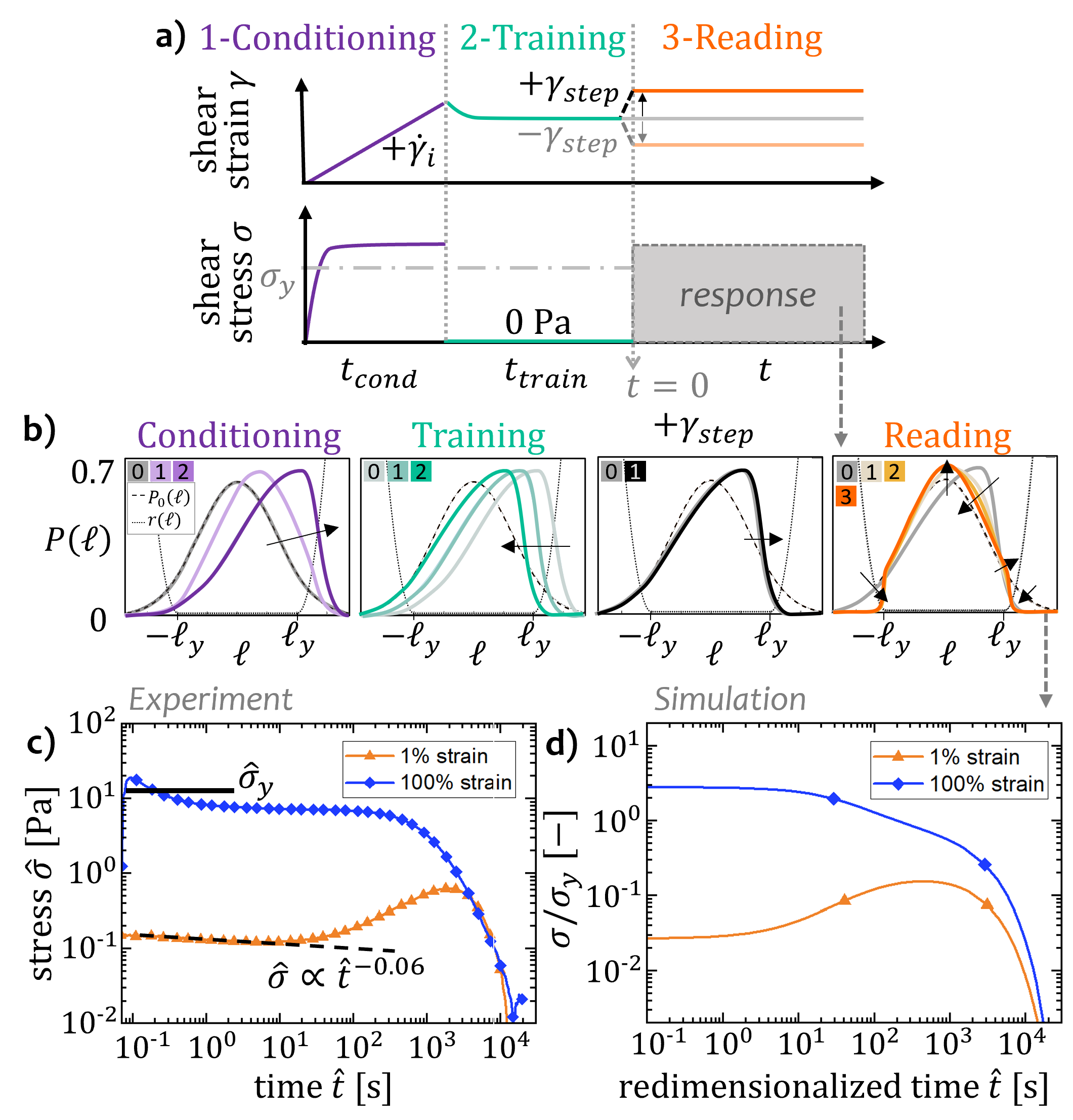}
\caption{\label{fig:protocol} 
(a) Residual Stress Probe protocol. (1) \textbf{Conditioning} at a high shear rate $\dot{\gamma}$ followed by  (2) \textbf{training} at zero stress for $t_{train}$ and a brief step strain as a probe, and (3) \textbf{reading} with shear rate held at zero. 
\textcolor{black}{(b) Material strain distribution $P(\ell)$ simulated with EP-PLY shown evolving per step from beginning (0) to end in spaced snapshots, for simulated  foam and 1\% step. 
(c) In \textbf{reading} of a young foam, the material shear stress $\hat{\sigma}$ relaxes slowly following a power law decay. 
(d) \textbf{Reading} simulated with EP-PLY presents constant stress at short times, rising for low strains, and decaying at longer times for all strains. 
Displayed axis scales are congruent between (c) and (d), with $\hat{\sigma}_y = 12$~Pa. 
}}
\end{figure}

\textcolor{black}{We concern ourselves with a new Residual Stress Probe protocol designed to elucidate precisely the influence of prior deformation on a still and stress-equilibrated material, and we decouple short-term and long-term relaxation into two separately tunable behaviors observed in non-thixotropic yield stress fluids (\textit{i.e.}, those without time-dependent viscosity). In particular, we carefully denote preshear as \textbf{conditioning}, an intermediary waiting at zero stress as \textbf{training}, and stress relaxation at fixed strain as \textbf{reading}, in analogy to memory experiments in solid materials \cite{PhysRevMaterials.6.L042601, PhysRevLett.123.218003, Murphy2020}, and we focus on the stress response of material in the final step (Fig~\ref{fig:protocol}). Experiments probe four jammed athermal soft materials, two Carbopol samples and a foam of two ages
\cite{owens2020improved, carraretto2022time}, using a shear rheometer.}

\textcolor{black}{
Our protocol generates two apparent classes of response: for probe strains high relative to the yield strain, in \textbf{reading} the stress relaxes slowly, and then quickly, while for low probe strains the stress increases at short times followed by a terminal relaxation at long times (shown for foam, Fig~\ref{fig:protocol}c). We recreate both behaviors with a strain distribution-based elastoplastic model for further insight (Fig~\ref{fig:protocol}b,d), and we find that Carbopol behaves analogously for the initial stress rise (Fig~\ref{fig:EM_Carbopol_alone}) with terminal relaxation outside of accessible measurement windows.}


\textcolor{black}{\textit{Background} -}  
{}{
The occurrence of two-part relaxation separated in timescales has been reported in simulations of soft glassy materials following stress cessation or jamming, which ascribe the phenomena to relaxation of particle clusters at early times and relaxation at later times of particle packings \cite{vinutha2023memory}, particle cage breaking and cage deformation \cite{khabaz2021transientPartI}, or localized plasticity compared to system-spanning effects \cite{barik2022origin}, while experimental XPCS describes a distribution of cluster relaxations \cite{song2022microscopic}. The feature itself relates well to existing toy models that have been used to capture the Kovacs effect using distributed relaxation of fast and slow elements \cite{Murphy2020}, though the prior model fails if a \textbf{training} step is included. 
Meanwhile, yielding of yield stress fluids has been shown to be a continuous rather than an instantaneous process, in which materials accumulate plastic strain at the microstructural level long before a yielding event occurs \cite{kamani2021unification}, supporting our findings of coherent sub-yield stress responses.}



\textit{Model Development} -  
{}{As our experimental results are observed for non-thixotropic fluids, we must find a model that }
{}{does not invoke thixotropy or shear-banding, which are typically implicated in time-dependent rheological complexities, and yet is sufficient to recreate the experimentally observed stress rise (Figs~\ref{fig:protocol}c,d).} \textcolor{black}{We therefore turn to elastoplastic models as a well-established method of continuum modeling for these kinds of materials \cite{Nicolas2018}. In particular, we construct a model that} {}{relies on the evolution of strain distributions, emphasizing strain elements and their direct contributions to material behavior. In contrast, the well-known Soft Glassy Rheology model uses an energy landscape with a distribution of trap depths and thermal activation to describe yielding and flow, and recreates a wide variety of aging effects \cite{fielding2000aging}.} \textcolor{black}{In our approach, we carefully modify yielding rules from existing elastoplastic models \cite{Barlow2020} to generate quantitative agreement with transient and steady state experiments, showing this choice of model to be minimal, sufficient, computationally tractable, and insightful.
}

Our mesoscopic elastoplastic model with power-law yielding (EP-PLY) is based on a time-varying distribution $P(\ell, t)$ of strain elements $\ell$ each corresponding to a local mesoscopic region of material \cite{Nicolas2018}. 
The governing equation for the distribution $P(\ell, t)$ is  
\begin{eqnarray}\label{eq:main}
\partial_t P(\ell,t) +\dot{\gamma}\partial_\ell P(\ell,t) = -r(\ell)P(\ell,t) + P_0(\ell)Y(t).
\end{eqnarray}

where $\partial_t$ and $\partial_\ell$ are partial derivatives in time and elemental strain, respectively. The distribution $P(\ell, t)$ is normalized to have an area of unity. 
The shear rate is $\dot{\gamma}$. The attempt rate for yielding $r(\ell)$ is  defined uniquely here as 
\begin{eqnarray} 
r(\ell) = \varepsilon + \left\{
\begin{array}{lr}
(-\ell/\ell_y - 1)^\nu &\ell/\ell_y < -1\\
0 & -1\leq\ell/\ell_y\leq1\\
(\ell/\ell_y - 1)^\nu & \ell/\ell_y > 1
\end{array}\right.
\label{eq:well}
\end{eqnarray}
which is parametrized by the yield strain $\ell_y$, the power-law well incline $\nu$, and the background fluidity $\varepsilon$. Notably $r(\ell)$ is the same for all states, rather than having a distribution of energies. The function is piecewise continuous with a continuous derivative for $\nu>1$ (our regime of choice), and reduces to the common ``top hat" shape for $\nu= \varepsilon =0$. 
Upon yielding, elements reset to positions within the Gaussian post-hop distribution $P_0(\ell)$ with width set by $\ell_0$: $P_0(\ell) = \exp{(\frac{1}{2}(\frac{\ell}{\ell_0})^2)}/\left(\ell_0\sqrt{2\pi}\right)$. The yielding rate $Y(t)$ connects the attempt rate and distribution $Y(t) = \int_{-\infty}^{\infty}{d\ell \, r(\ell)P(\ell, t)}$. 

Thus, the four terms of Equation~\ref{eq:main} incorporate, from left to right, the effects of time evolution, elastic loading in shear, plastic yielding, and rebirth events. 
As is typically used in this genre of model, the total elastoplastic stress is the first moment of the distribution, $\sigma(t) = \int_{-\infty}^{\infty}{d\ell\, \ell\, P(\ell, t)}$ \cite{Nicolas2018}. \textcolor{black}{The evolution of $P(\ell)$ within our full residual stress probe protocol is depicted in Fig~\ref{fig:protocol}b.}



\begin{figure}
\includegraphics[width=0.92\columnwidth]{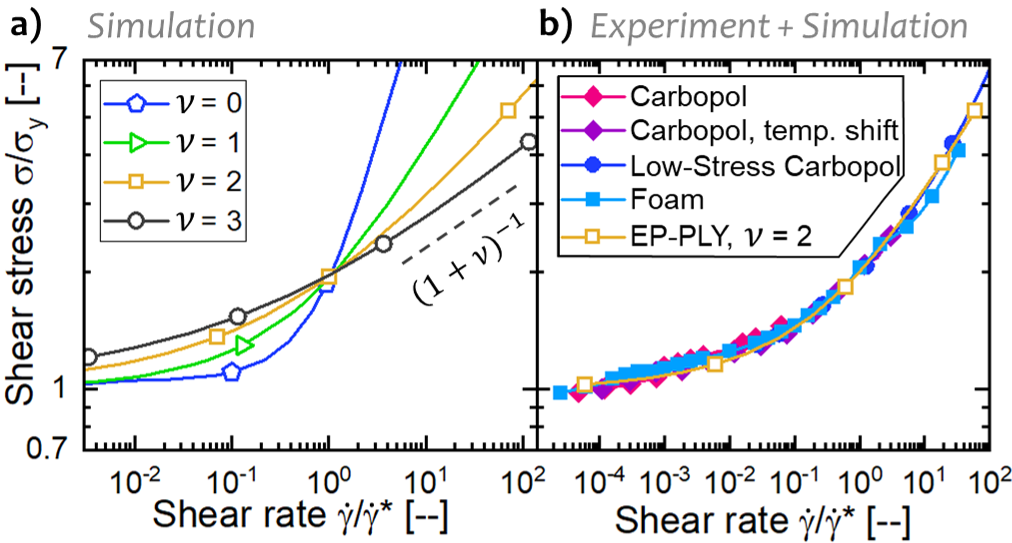}
\caption{\label{fig:flow_curves} \textcolor{black}{Tuning the steady shear flow curve. 
(a) Flow curves simulated by EP-PLY vary only the yielding rate power laws $\nu$, with high-$\dot{\gamma}$ power laws of $n=(1+\nu)^{-1}$. 
(b) Experimental flow curves from one Carbopol solution at room temperature and with time-temperature superposition to extend the measurement window, the same Carbopol solution having been vigorously mixed, Gillette Original Foam, and a simulated flow curve for $\nu = 2$, all collapsed by $\dot\gamma^*$ and $\sigma_y$.}
}
\end{figure}

In order to compare between simulation and experiments, we selected strain, time, and stress scales. For clarity, we denote with a hat $(\hat{\cdot})$ symbols with dimension whereas bare units $(\cdot)$ are dimensionless. 
The macroscopic strain $\gamma$ is based on elemental strain $\ell$. We observe that the steady-state flow curves from EP-PLY (Fig~\ref{fig:flow_curves}a) \textcolor{black}{have an apparent Herschel-Bulkley (HB) form, $\sigma(\dot{\gamma})=\sigma_y + k\dot{\gamma}^n$, for yield stress $\sigma_y$, consistency index $k$, and shear thinning index $n$. We define a characteristic HB shear rate as $\hat{\dot{\gamma}}^* \equiv (\hat{\sigma_y}/\hat{k})^{(1/n)}$ and select our timescale as its inverse $\hat{\tau}_0 \equiv 1/\hat{\dot{\gamma}}^*$. We take the yield stress $\hat{\sigma}_y$ as our stress scale.}
Using these scalings, flow curves between experiments and simulation exhibit qualitative and quantitative similarity for our materials, as the well power law $\nu$ controls the power law in the flow curve $n$ (Fig~\ref{fig:flow_curves}b). 
\textcolor{black}{As an aside, in the limit of $\ell_0=0$ and $\varepsilon=0$ (used for solvability), EP-PLY analytically resolves to a flow curve of three parts, rather than the two of HB, similar to proposed alternatives \cite{caggioni2020variations} (Sec~\ref{SI:3curve}).} 

To complete our model-experiment link, values for the post-hop distribution width $\ell_0/\ell_y$ and background fluidity $\varepsilon$ were selected to correspond to observed experimental stress relaxation results. The values for $\ell_0/\ell_y$ ranged from 0.35 (aged foam) to 0.80 (young foam), with intermediate values for Carbopol (0.65 for the standard sample, 0.50 for the highly-mixed sample). These values all lie within the relatively ductile regime for EP models \cite{Barlow2020} and the lower values for older foam align with results showing that soft glasses may embrittle with age \cite{PetekidisTwoStepYielding}.  
Full test parameters are detailed in Sec~\ref{SI:modelparams} \textcolor{black}{with additional exploration of model parameters in Sec~\ref{SI:model_behavior}.}



\begin{figure}
\includegraphics[width=0.93\columnwidth]{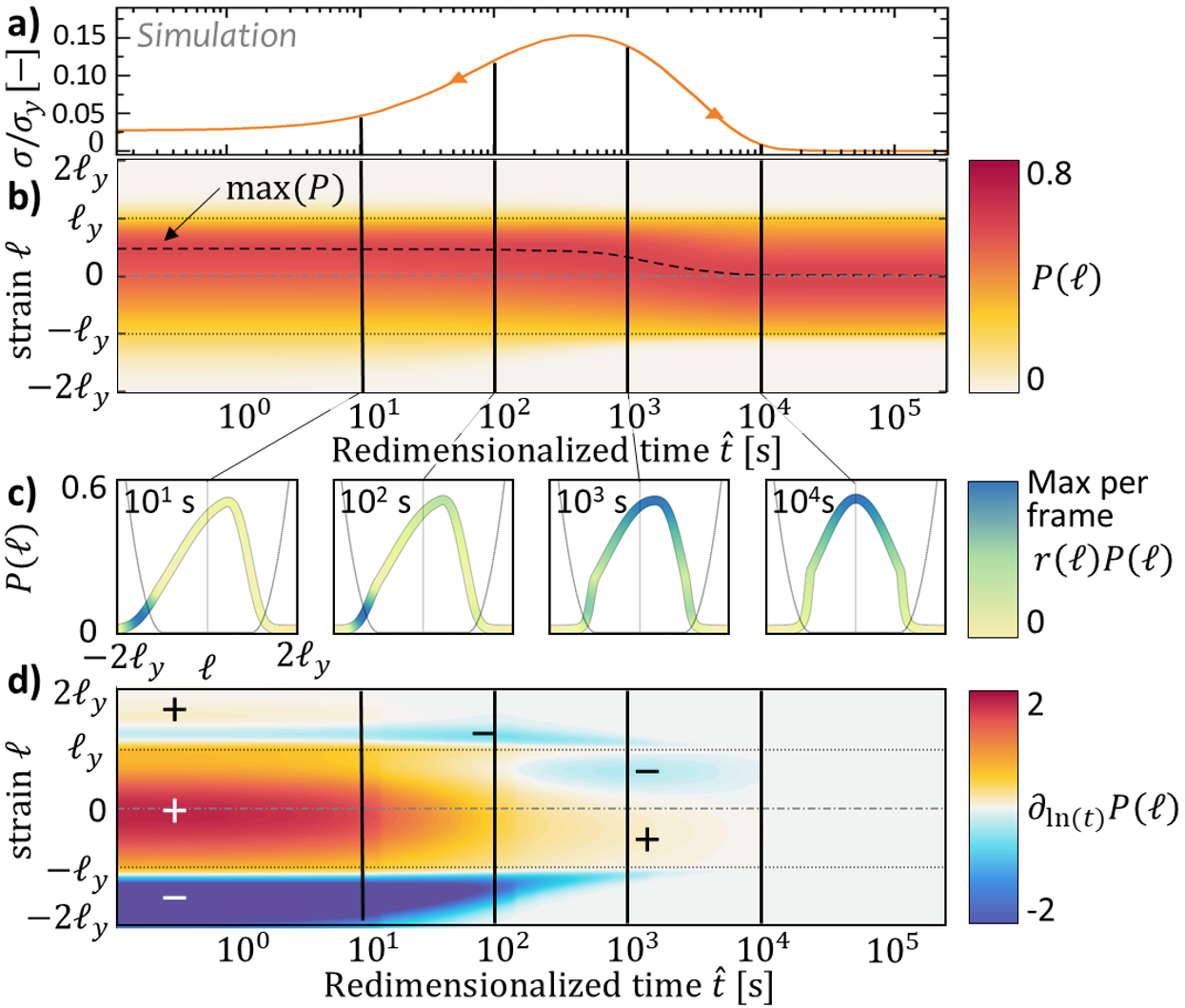}
\caption{\label{fig:kymograph} 
Simulated response of foam for probe strain $\gamma_{step} = 0.01$, showing 
(a) the initial stress response from Fig~\ref{fig:protocol}d. 
(b) A time-strain plot shows simulated values for the distribution $P(\ell)$ over time. Color indicates magnitude of $P(\ell)$ and the dashed line tracks the distribution maximum, from slightly positive at early times to centered on $\ell = 0$ at late times. 
(c) Strain element distributions, $P(\ell)$, predicted by EP-PLY during relaxation. Curves are colored by the yielding rate $Y(\ell) = r(\ell)P(\ell,t)$ per element $\ell$, showing anisotropy in $Y(\ell)$ at multiple times. \textcolor{black}{The color scale is normalized per subpanel to focus on locations of highest relative yielding.}
(d) A time-strain plot shows the change in the element distribution over time, $\partial_{ln(t)}P$. Early times show a strong decrease at large negative strains $\ell<-\ell_y$, and late times primarily show a decrease at small positive strains $0<\ell<\ell_y$, indicated by $``+"$ and $``-"$ markers.
}
\end{figure}

\begin{figure*}
\includegraphics[width=0.89\textwidth]{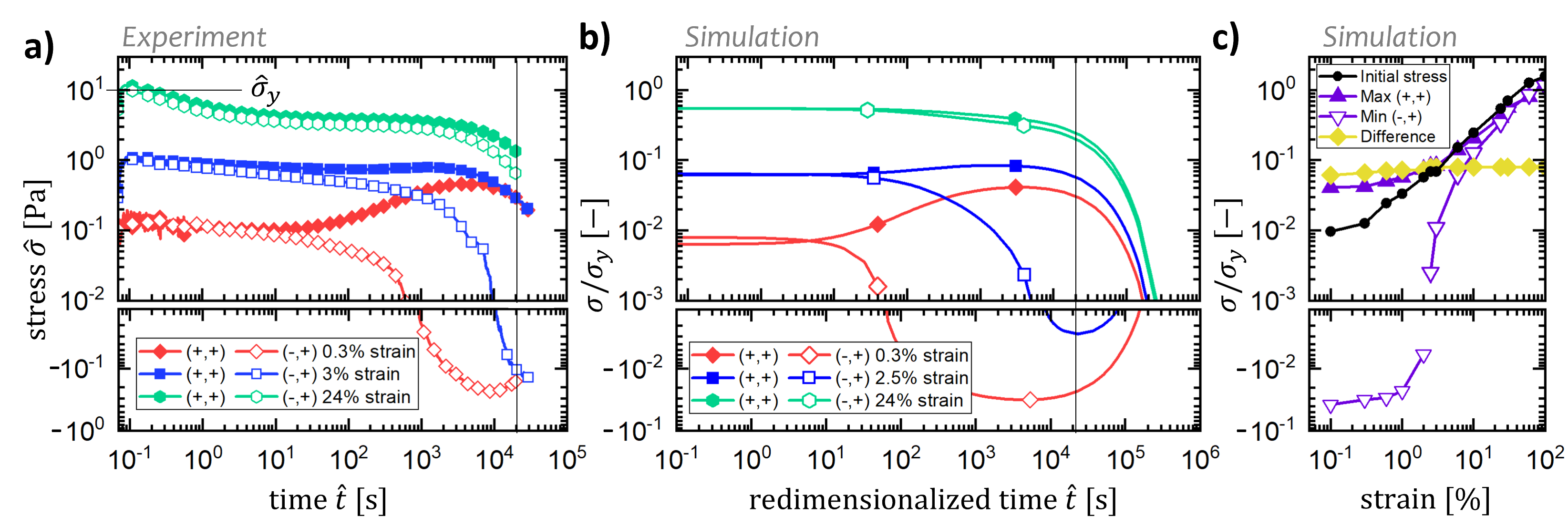}
\caption{\label{fig:directionality} Influence of directionality of \textbf{conditioning}. 
(a) Experimental response of old foam to stress relaxation after step strains of $\gamma_{step} = 0.3\%$,  $3\%$, and $24\%$. Steps follow conditioning in the same direction, filled symbols and $(+,+)$, or reversed direction, hollow symbols and $(-,+)$. A vertical bar at $3\times10^4$~s indicates the practical time limit of experiments. 
(b) Simulated response of foam-like material in simulation space. 
Displayed axis scales are congruent between (a) and (b), with $\hat{\sigma}_y = 10$~Pa.
(c) Results from a range of experiments with $\gamma_{step} = 0.1$-$100\%$ show the initial stress at the start of \textbf{reading}, absolute stress after about $10^3$~s in (b) for both step directions, and the difference between stress maximum in $(+,+)$ and minimum in $(-,+)$. 
}
\end{figure*}

\textit{Distribution of Local Strain} - 
EP-PLY makes a prediction of the internal mechanisms that may lead to an increase in stress over long times when strain is held fixed at zero. We focus on the evolution of the shape of the distribution function $P(\ell)$ over time. In short, the distribution of strain states shows two discrete phenomena occurring on two separate timescales leading to the rise, and later fall, of stress during relaxation (Fig~\ref{fig:kymograph}a). 

This is visualized by plotting the distribution $P(\ell)$ as a continuous kymograph \cite{aside} (Fig~\ref{fig:kymograph}b) and at select times (Fig~\ref{fig:kymograph}c). At the start of \textbf{reading}, $P(\ell)$ retains a slightly longer negative tail at large strains $\ell<-\ell_y$ and a small hump at small positive strains $0<\ell<\ell_y$. \textcolor{black}{While EP-PLY imposes no size scale, as a rough mental picture one could imagine the material consisting of some particles/clusters/domains very stretched in a negative shear strain direction (opposing the directions of \textbf{conditioning} and the step), balanced by most particles/clusters/domains stretched slightly in the positive direction.} Initially, this long negative tail is subjected to a higher yielding rate $Y(\ell)$ and yields relatively quickly, repopulating with a centered distribution according to $P_0(\ell)$ and causing the stress ($\int_{-\infty}^{\infty}{d\ell\,\ell\,P(\ell, t)}$) to increase by removing the most negative portion. At later times set by $\varepsilon$, the small positive hump subject to a lower yielding rate $Y(\ell)$ finally yields, relaxing stress entirely. 
A kymograph of $\partial P/\partial \ln t$, the change over log time, shows points of variation in distribution shape over time more distinctly (Fig~\ref{fig:kymograph}d). 

Our results indicate that the nonmonotonic stress response observed in the model is due to the small amount of incomplete relaxation during \textbf{training}, which becomes embedded in a biased distribution of microstructural strain states
\textcolor{black}{that then yield at different rates based on their locations within the distribution.}
The time required to fully relax all initial stresses during \textbf{training} is beyond achievable experimental windows, interfered with by 
practical obstacles
in experimental conditions above 10$^5$ seconds. One real-world experiment was conducted with a foam sample created \textit{in situ} in the rheometer, which was immediately subjected to \textbf{reading} alone. In this case, the stress rise was an order of magnitude lower than for one following our full residual stress probe protocol (Sec~\ref{SI:no_preshear}, Fig~\ref{fig:SI-foamed-in-place}). 
\textcolor{black}{In simulation, a fully relaxed state prior to \textbf{reading} presents a response that decays monotonically to zero stress
(Fig~\ref{fig:SI_directionality_logspace}).}


\textit{Directional Memory} -  
Building on our understanding of how the \textbf{reading} response is likely embedded, we \textcolor{black}{consider} the \textcolor{black}{stress growth and} non-monotonic stress  response \textcolor{black}{already presented, and directly relate it} to \textbf{conditioning} by varying the direction of the pre-shear, but maintain the probe step direction (Fig~\ref{fig:protocol}a). A notation of $(+,+)$ indicates \textbf{conditioning} in the positive \textcolor{black}{strain} (counterclockwise) direction followed by training at zero stress and then a step strain in the positive (ccw) direction. In contrast, a notation of $(-,+)$ indicates \textbf{conditioning} in the reverse (cw) direction, \textcolor{black}{which createas a long positive tail in $P(\ell)$, causing stress relaxation to occur in the negative stress direction}. After application of $(+,+)$ and $(-,+)$ for a step in strain low, medium, and high relative to the material yield strain, directional memory retention was seen in experiments and simulations (Fig~\ref{fig:directionality}a,b). 
For small steps, a split-direction $(-,+)$ test yielded a response dominated by residuals, passing from positive down through negative values and ending at zero, while for large steps the apparent difference was small compared to the overall stress response, and nearly invisible on a log scale.
\textcolor{black}{This serves to generalize our earlier results by showing how the underlying phenomena leading to anomalous stress rise can be reversed in direction and can additively (subtractively) influence tests with larger step strain, even if the full response is monotonic.}
Results for Carbopol are analogous  (Fig~\ref{fig:SI_directionality_logspace}).

A larger set of strain steps were simulated, spanning 0.1\%-100\%. For the initial (primarily elastic) stress response at early times ($\hat{t}<1$~s), the stress increased approximately as a simple power law of the applied strain. At later times ($\hat{t}>10^3$ s), the maximum absolute stresses during \textbf{reading} increased with strain, due to the larger elastic response. However, the difference in $(+,+)$ and $(-,+)$ steps increased only marginally over this same range, remaining perceptible over the full range of tested strains, and retaining a value slightly less than one tenth of the yield stress $\sigma_y$ (Fig~\ref{fig:directionality}c), a not-insignificant value, \textcolor{black}{\textit{even when the initial stress response exceeded the yield stress} (Fig~\ref{fig:directionality}a,c)}.
\textcolor{black}{This result indicates that the influence of \textbf{conditioning} on  internal strain distribution persists in superposition to even large elastic strains and remains present even at long times ($>10^4$~s here). Thus, the influence of prior deformation history in existing literature is likely severely under-noticed. }

\textcolor{black}{This oversight occurs in part because our residual stress probe protocol poses several advantages for directly measuring residual stress, which is often more subtle.}
The long-time terminal relaxation and viscoelatic response of our identical foam has been examined using extended creep \cite{Lavergne2022}. Compared to our protocol, creep is sustained \textbf{training}, and did not prompt a non-monotonic response even for low rates \cite{Lavergne2022}, nor for similar tests in Carbopol \cite{nikoumanesh2024elucidating}, showing the need for \textbf{reading} via stress relaxation in order to uncover the phenomena we report. 

\textcolor{black}{In addition, the response here is phenomenologically reminiscent of non-monotonic stress relaxation observed in rheological measurements following flow cessation in associative supramolecular gels \cite{PhysRevLett.123.218003}, Boehmite gels \cite{PhysRevMaterials.6.L042601}, and simulated shear-banding fluids \cite{ward2024shear}, linked to associativity, overaging, and shear bands, and bears similarity to the Kovacs effect in heterogeneous aggregates observed at the moment compression is partially released \cite{RevModPhys.91.035002, Murphy2020, kursten2017giant}. In contrast, 
stress relaxation here follows an intermediary zero-stress step, allowing cessation of (most) motion and equilibration of all bulk stress, which provides a more direct measurement. This is also in contrast to other studies that employ indicators of residual stress as small features overlaid on a main experiment (\textit{e.g.}, \cite{edera2025mechanical, Choi2020}). Second, the ability to alter the magnitude of the step strain here allows fine-tuning of the expected initial elastic response in proportion to residual stress, which is directly responsible for our results in Figure~\ref{fig:directionality}c that demonstrate how even large elastic responses do not eliminate the residuals, but in superposition merely overshadow them.}

\textcolor{black}{Our results raise the question of whether an alternative to high-shear preconditioning can better reset materials to a ``zero" state. One promising suggestion is a period of low-level oscillations \cite{edera2025mechanical}.  According to EP-PLY, oscillations of an appropriate magnitude would systematically erode long tails of $P(\ell)$. This may also allow controlled mechanical annealing of soft materials, as also suggested \cite{edera2025mechanical} by sharpening $P(\ell)$ below the width set by the material's native $\ell_0$.}

\textcolor{black}{Finally, our results probe at whether yield stress materials ultimately relax if given unreasonably long time.}
We remark that a non-zero background fluidity $\varepsilon$ is necessary to recreate our experimentally observed stress relaxation, and this also causes the EP-PLY steady state flow curve to intercept $\sigma = 0$ for $\dot{\gamma} = 0$.
An absence of yield stress at \textcolor{black}{impractically} low shear rates has been suggested based on experimental evidence \cite{barnes1985yield, barnes2007yield, younes2021situ} although \textcolor{black}{measurement artifacts like slip are difficult to exclude in this low-$\dot{\gamma}$ range. Through basic analysis, we find that our results indicate finite zero-shear viscosities $\eta_0$ for the materials tested (Sec~\ref{EM:eta0}.)}
A finite $\eta_0$ would indicate a lack of ``true" yield stress for these materials, so our estimates would need to be confirmed by further tailored study. 

\textcolor{black}{Overall, we have proposed a new experimental protocol to probe residual stress in yield stress fluids apparent after past deformation. We have constructed an elastoplastic model, EP-PLY, to explain how this arises---from an asymmetric distribution of internal strain states that are able to store long-term memory---thus showing time-dependent behavior without thixotropy or shear-banding. EP-PLY further agrees quantitatively with experimental steady flow curves, suggesting its usefulness more widely as a material model.}  
\\

We thank Suzanne Fielding and Gareth McKinley for a series of foundational discussions about our model and design of suitable experimental analysis. We also thank Yogesh Joshi, Gavin Donley, Simon Rogers, Emanuela Del Gado, and H. A. Vinutha for discussions of various results; and we kindly acknowledge the Mathworks Engineering Fellowship from MIT Department of Mechanical Engineering and the Postdoctoral Fellowship Program for Engineering Excellence from the MIT School of Engineering for financial support.




\begin{thebibliography}{31}
\bibitem [{asi()}]{aside}%
  \BibitemOpen
  \href@noop {} {\ }\bibinfo {note} {We use `Kymograph' to denote a plot
  showing the change in a function over time, although it is not a `space' plot
  as the name would suggest, but rather is a function of strain element. For
  our use it is interchangeable with a time-strain plot.}\BibitemShut {Stop}%
\end{thebibliography}
\providecommand{\noopsort}[1]{}\providecommand{\singleletter}[1]{#1}%
%



\pagebreak

\renewcommand{\thefigure}{EM-\arabic{figure}}
\setcounter{figure}{0}

\renewcommand{\thesection}{EM-\arabic{section}}
\setcounter{section}{0}

\renewcommand{\theequation}{EM-\arabic{equation}}
\setcounter{equation}{0}

\clearpage 

\section{End Matter}

\subsection{Analytical yield stress from model} \label{SI:analytical_yield_stress}

Equation~\ref{eq:main} was solved analytically for the value of the yield stress for $\varepsilon=0$, giving
\begin{equation}
    \sigma_y = \frac{1}{2}\left(1-\ell_0^2\right) + \frac{\ell_0 e^{(-1/2\ell_0^2)}}{\sqrt{2\pi}\erf(\frac{1}{\ell_0\sqrt{2}})}
\end{equation}
In the limiting case of $\ell_0\rightarrow0$, $\sigma_y=1/2$ and for $\ell_0\rightarrow\infty$, $\sigma_y=1/3$. This holds for arbitrary $\nu$.

\subsection{Analytical three-component flow curve} \label{SI:3curve}

\begin{figure}
\includegraphics[width=0.75\columnwidth]{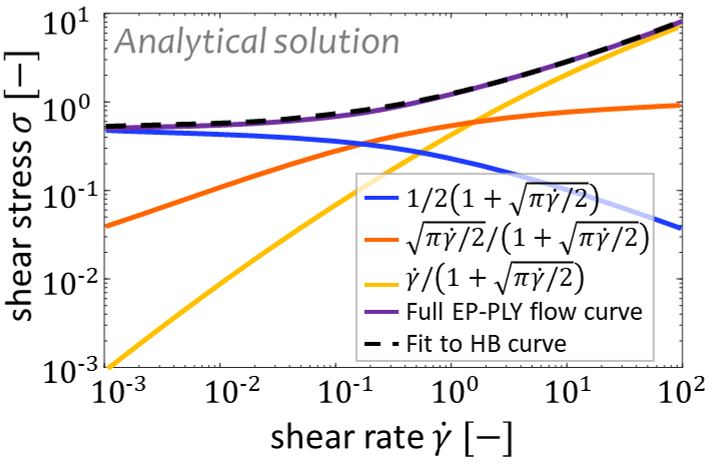}
\caption{\label{fig:multipart_flow_curve} 
The three sub-parts of the analytical flow curve and their sum, compared with a Herschel-Bulkley (`HB') fit. 
}
\end{figure}

We solved for an analytical solution to the EP-PLY model, Equation~\ref{eq:main}, in the limits of $\ell_0 \rightarrow 0$ and $\varepsilon=0$, for the case when $\nu>1$,  
\begin{equation} \label{eq:3curve}
    \sigma = \frac{1/2 + \beta + \beta^2/(\nu+1)^2}{1 + \beta} 
\end{equation}
\begin{equation}
    \beta \equiv \left(\frac{\dot{\gamma}}{(\nu+1)^\nu}\right)^{1/(\nu+1)}
\end{equation}
Equation~\ref{eq:3curve} can be checked in the limiting cases. First, in the low-shear-rate limit, $\lim_{\dot{\gamma}\to0} \sigma = 1/2$
which gives a constant value and aligns with results from Section~\ref{SI:analytical_yield_stress} for the case taken here of $\ell_0=0$ and $\varepsilon=0$. In the high shear rate limit,  
\begin{equation}
    \lim_{\dot{\gamma}\to\infty} \sigma = \left[\frac{1}{(\nu+1)^{\nu-2}}\right]^{1/(1+\nu)}\dot{\gamma}^{1/(1+\nu)}
\end{equation}
which shows constant flow, and allows direct selection of $\nu$ to represent a real material by comparison with the shear flow term in Herschel-Bulkley flows, $\lim_{\dot{\gamma}\to\infty} \sigma_{HB} = k\dot{\gamma}^n$. Thus, $n=1/(1+\nu)$ or $\nu=(1/n)-1$ for $\nu\geq1$. 
The only free parameter is $n$, or equivalently $\nu$, which fully determines the stress behavior as a function of shear rate. For the more general case with $\ell_0>0$, it is expected that the expression would vary, \textit{i.e.}, $\sigma = f(\nu, \ell_0, \dot{\gamma})$. 
The equation follows a similar form for $\nu=1$,  
\begin{equation} \label{eq:nu1}
    \sigma = \frac{1/2 + \sqrt{\pi\dot{\gamma}/2} + \dot{\gamma}}{1 + \sqrt{\pi\dot{\gamma}/2}}. 
\end{equation}
The three sub-components and their sum are plotted in Fig~\ref{fig:multipart_flow_curve} along with a Herschel-Bulkley fit to the flow curve, showing near-perfect agreement in the shape of our function derived from a mesoscopic elastoplastic model and the long-standing empirical model. 

The flow curve of Equation~\ref{eq:3curve} is similar to a multipart flow curve previously proposed \cite{caggioni2020variations} based on experimental evidence that the shear stress, $\sigma$, is 
\begin{equation}
    \sigma = \sigma_y +  \sigma_y \cdot \left(\frac{\dot{\gamma}}{\dot{\gamma}_c}\right)^{1/2} + \eta_{bg}\cdot\dot{\gamma}
\end{equation}
which decomposes to, from left to right, elastic, plastic, and viscous contributions to stress where $\dot{\gamma}_c$ is a critical shear rate and $\eta_{bg}$ is the background/solvent viscosity. In our case of Equation~\ref{eq:3curve} or Equation~\ref{eq:nu1}, the leftmost term arises from the stress contribution of elements within the well, $-\ell_y<\ell<\ell_y$, which are by nature unyielding, corresponding to elastic effects. The second two terms come from integration of elements outside the well, $\ell\leq-\ell_y$ and $\ell\geq\ell_y$, which correspond to a mix of plastic and viscous flow due to element yielding events, and are strongly affected by the shape of $r(\ell)$. 

\subsection{Test without \textbf{conditioning}} \label{SI:no_preshear}

\begin{figure}[h!]
\includegraphics[width=0.75\columnwidth]{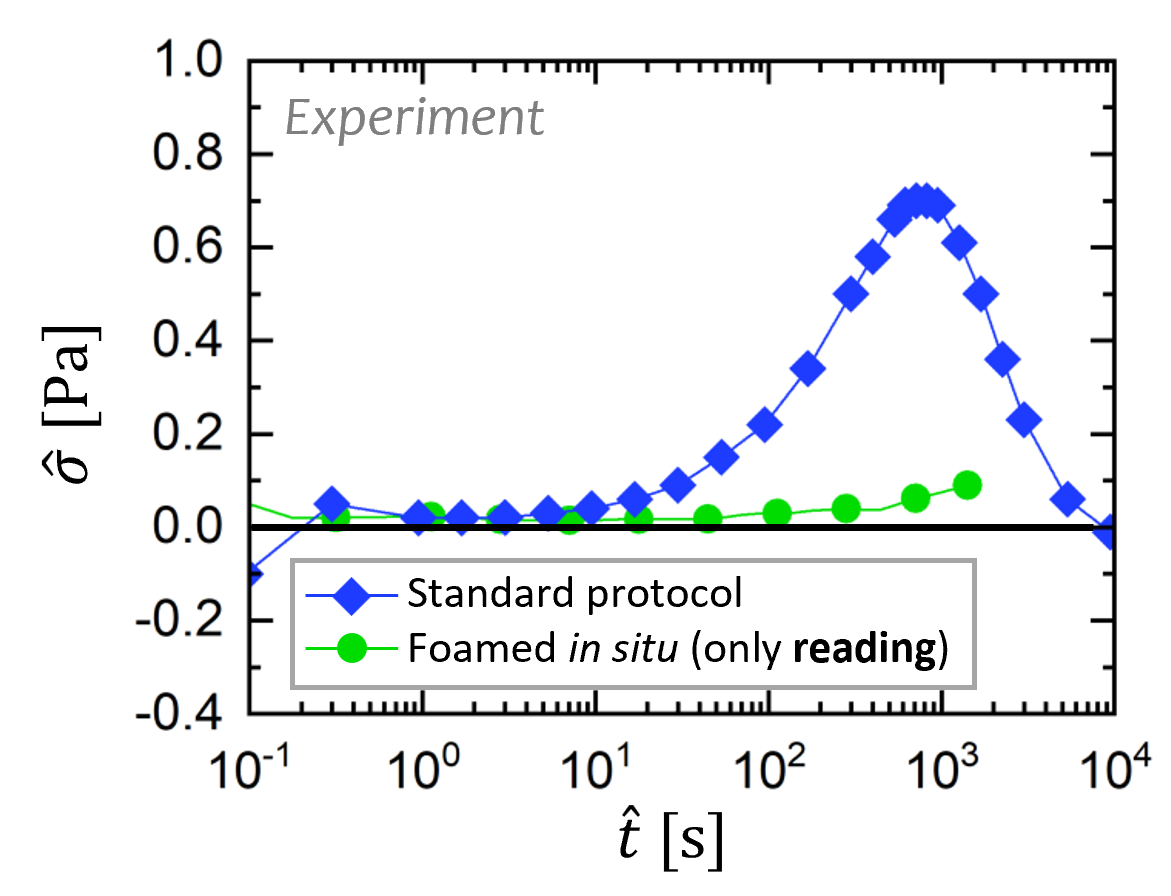}
\caption{\label{fig:SI-foamed-in-place} 
Two experiments on a young foam are compared, one using our standard protocol and one using a foam sample created \textit{in situ} and subjected to \textbf{reading} alone. 
}
\end{figure}

Two experiments are compared, one using a standard residual stress probe protocol (Fig~\ref{fig:protocol}) and one using a foam sample created \textit{in situ} and immediately subjected to \textbf{reading} alone (Fig~\ref{fig:SI-foamed-in-place}), both at very small probe strain 0.0035\%.
In the normal protocol, the sample showed a stress rise and fall. For the \textit{in situ} foam, the stress rose an order of magnitude less, though at the same time point $10^1$~s. We ascribe this small rise to material shear from the foaming process itself. This foam has a yield stress of $\hat{\sigma}_y=12$~Pa. 

\subsection{Carbopol}

In response to step strain imposed on a Carbopol microgel, the material shear stress $\hat{\sigma}$ relaxes slowly following a power law decay, remaining monotonic for materials subjected to larger steps, $\gamma_{step}\geq2.4\%$. After small steps, $\gamma_{step}<2.4\%$, the shear stress evolves non-monotonically, increasing in time up to levels of 20\% of the material yield stress ($\hat{\sigma}_y = 10$~Pa). Material simulated with EP-PLY presents constant stress at short times, decaying at longer times after higher probe strains, $\gamma_{step}\geq10\%$, and increasing after low probe strains, $\gamma_{step}<10\%$.

\begin{figure}
\includegraphics[width=1\columnwidth]{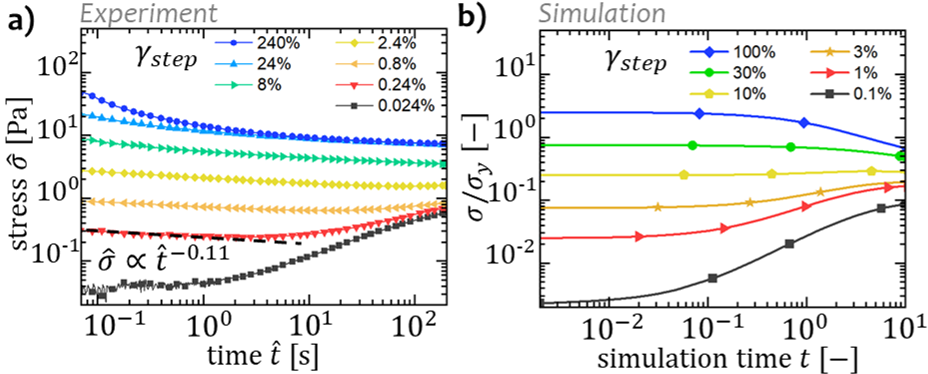}
\caption{\label{fig:EM_Carbopol_alone} Response to \textbf{reading} in a Carbopol microgel (a) in experiments and 
(b) as simulated with EP-PLY.
}
\end{figure}

\subsection{Simulated and experimental strain}

\begin{figure}
\includegraphics[width=0.95\columnwidth]{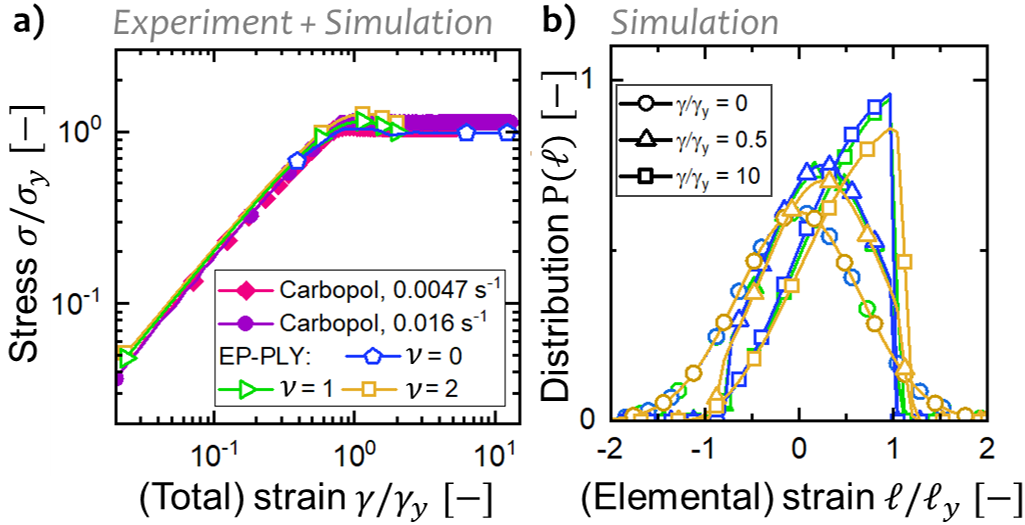}
\caption{\label{fig:SI_model-vs-experiments} Comparison of basic material properties via experimental and simulated values. 
(a) Start-up of steady shear rate test showing elastic-to-plastic transition of material. Simulations were performed for a material with post-hop distribution width of $\ell_0/\ell_y = 0.65$, background fluidity $\varepsilon = 0$, and yielding well exponents $\nu = 1, 2, 3$.  
(b) Simulated distributions at early, moderate, and late strains in (a) for each exponent case. 
}
\end{figure}

The macroscopic strain scale $\gamma$ used to compare between experiments and simulations is based on strain on elements $\ell$, as determined by simulated start-up of steady shear tests (Fig~\ref{fig:SI_model-vs-experiments}a), which showe initial elastic response followed by plastic yielding. In this kind of experiment, thixotropy is often seen as an overshoot in the increasing stress before decreasing to the final plateau value, or else decreasing continuously. The experimental and simulation results both show small overshoots, up to $3\%$ above the yield stress for Carbopol and up to $11\%$ for the simulation at $\nu=2$, with decreasing values for decreasing $\nu$. 
These are considered to be negligible here. 
The plateauing behavior of the stress response in EP-PLY simulation is due to the distribution reaching a steady yielding rate from the shape of the attempt rate well, which begins to rise rapidly for $\ell/\ell_y>1$ (Fig~\ref{fig:SI_model-vs-experiments}b).  



\subsection{Zero-shear viscosity $\eta_0$}\label{EM:eta0}
To further consider behavior at low rate, we examined the zero-shear viscosity $\eta_0$, which can be determined by integration of stress relaxation in response to a step $\gamma_{step}$ as  $\eta_0 = \int_{t=0}^\infty{\sigma(\tau)/\gamma_{step}\,d\tau}$.
Applied to data (Fig~\ref{fig:protocol}c), this evaluates to $\eta_0=2$~kPa.s.
In a second approach, a functional form of the flow curve is fit to the EP-PLY model outputs
as $\sigma = (\sigma_y + k\dot{\gamma}^n)/(1+\varepsilon/2\dot{\gamma})$. This equation is used directly to compute $\eta_0 = \lim_{\dot{\gamma}\to0} \sigma/\dot{\gamma} = 2\sigma_y/\varepsilon=3$~kPa.s. In comparison, the value for Carbopol is larger, estimated to be $\eta_0=40$ and 600 kPa.s by the respective methods.
A finite $\eta_0$ would indicate a lack of ``true" yield stress for these materials, so our estimates would need to be confirmed by further tailored study. \textcolor{black}{However, adding minor complexities to $\varepsilon$ such as dependence on shear rate could both preserve the final stress release and an apparent yield stress, although we do not explore such variations in the fluidity expression here.}

\newpage{}
\newpage
\pagebreak [4]
\pagebreak

\renewcommand{\thefigure}{SI-\arabic{figure}}
\setcounter{figure}{0}

\renewcommand{\theequation}{SI-\arabic{equation}}
\setcounter{equation}{0}

\renewcommand{\thesection}{SI-\arabic{section}}
\setcounter{section}{0}

\section{Supporting Information}

\subsection{Experimental materials and rheometer tests} \label{SI:materials}

The materials include a commercial Carbopol solution as used in \cite{owens2020improved}, the same Carbopol having been ultrasonicated to reduce the yield stress by a factor of 10, and a foam (Gillette Original) at young ($10^3$~s) and old ($10^5$~s) times. Experiments were performed using a textured cone-and-plate and a fractal vane geometry in a textured cup on both a stress-controlled DHR-3 rheometer and a strain-controlled ARES rheometer, producing equivalent results. Results are shown for data taken on a DHR with a cone and plate roughened with adhesive sandpaper for Carbopol and a vane for the more delicate foam. 
While the foam samples do undergo aging, the effect is small or negligible over the timescale of our experiments \cite{carraretto2022time}. 
All four resulting samples retain so-called ``simple" yield stress behavior, but vary in their yield stress and material timescales.


\subsection{Exploration of Model Behavior}\label{SI:model_behavior}

\begin{figure*}
\includegraphics[width=110mm]{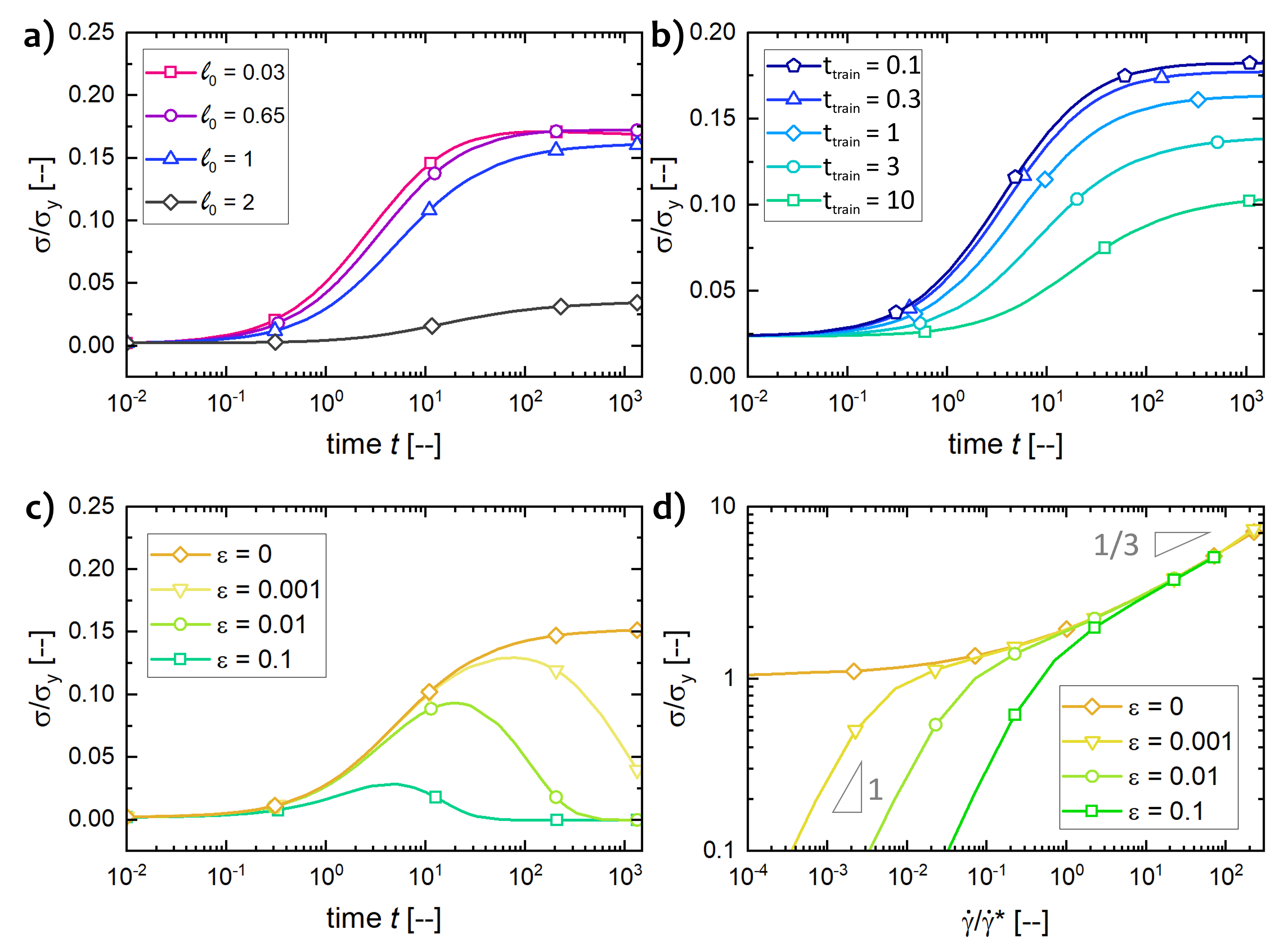}
\caption{\label{fig:SI_simulation-sweep} (a) Stress, $\sigma/\sigma_y$, over time in the \textbf{reading} step for a range of tested post-hop distribution widths, $\ell_0/\ell_y=0.03$ to $2$. Simulations were performed for a material with background fluidity $\varepsilon = 0$, and yielding well exponent $\nu = 2$.  
(b) Stress over time for variations in the duration of training shown for $t_{train} = 0.1, 0.3, 1, 3$ and $10$. Simulation parameters are the same as in (a) with $\ell_0/\ell_y = 0.65$.
(c) Stress over time in the \textbf{reading} step for a fixed value of the post-hop distribution widths, $\ell_0/\ell_y = 0.65$ and varying the background fluidity $\varepsilon = 0, 10^{-3}, 10^{-2}$ and $10^{-1}$. 
(d) Steady-state rheological flow curves emerging from the same parameter choices as in (c), following a low-shear rate power law of $1$ when $\varepsilon > 0$ and high-shear rate power law of $1/3$.
}
\end{figure*}

We explore the model behavior in Fig~\ref{fig:SI_simulation-sweep}. First, the stress, $\sigma/\sigma_y$, is shown as a function of time in the \textbf{reading} step for a range of tested post-hop distribution widths, $\ell_0/\ell_y=0.03$ to $2$ (Fig~\ref{fig:SI_simulation-sweep}a). Next, stress over time is shown for variations in the duration of training shown for $t_{train} = 0.1, 0.3, 1, 3$ and $10$ (Fig~\ref{fig:SI_simulation-sweep}b). Finally, stress over time is shown in the \textbf{reading} step for a fixed value of the post-hop distribution widths, $\ell_0/\ell_y = 0.65$ and varying the background fluidity $\varepsilon = 0, 10^{-3}, 10^{-2}$ and $10^{-1}$ (Fig~\ref{fig:SI_simulation-sweep}c). Corresponding, steady-state rheological flow curves in Fig~\ref{fig:SI_simulation-sweep}d are shown using same parameter choices as in Fig~\ref{fig:SI_simulation-sweep}c, and exhibiting a low-shear rate power law of $1$ when \textcolor{black}{$\varepsilon > 0$} and high-shear rate power law of $1/3$.

As seen in Fig~\ref{fig:SI_simulation-sweep}d, a non-zero background fluidity $\varepsilon$ also causes the steady state flow curve to intercept $\sigma = 0$ for $\dot{\gamma} = 0$, and reduces the yield stress plateau in the flow curve to begin at approximately $\dot{\gamma} \gtrsim 30/\varepsilon$. 
This takes the functional form, found by fitting EP-PLY model outputs with zero error, as 
\begin{equation} \label{eq:flowcurve}
    \sigma = \frac{\sigma_y + k\dot{\gamma}^n}{1+\varepsilon/2\dot{\gamma}}. 
\end{equation}

\subsection{Long-term relaxation of Carbopol tested in different directions}

\begin{figure*}[b]
\includegraphics[width=145mm]{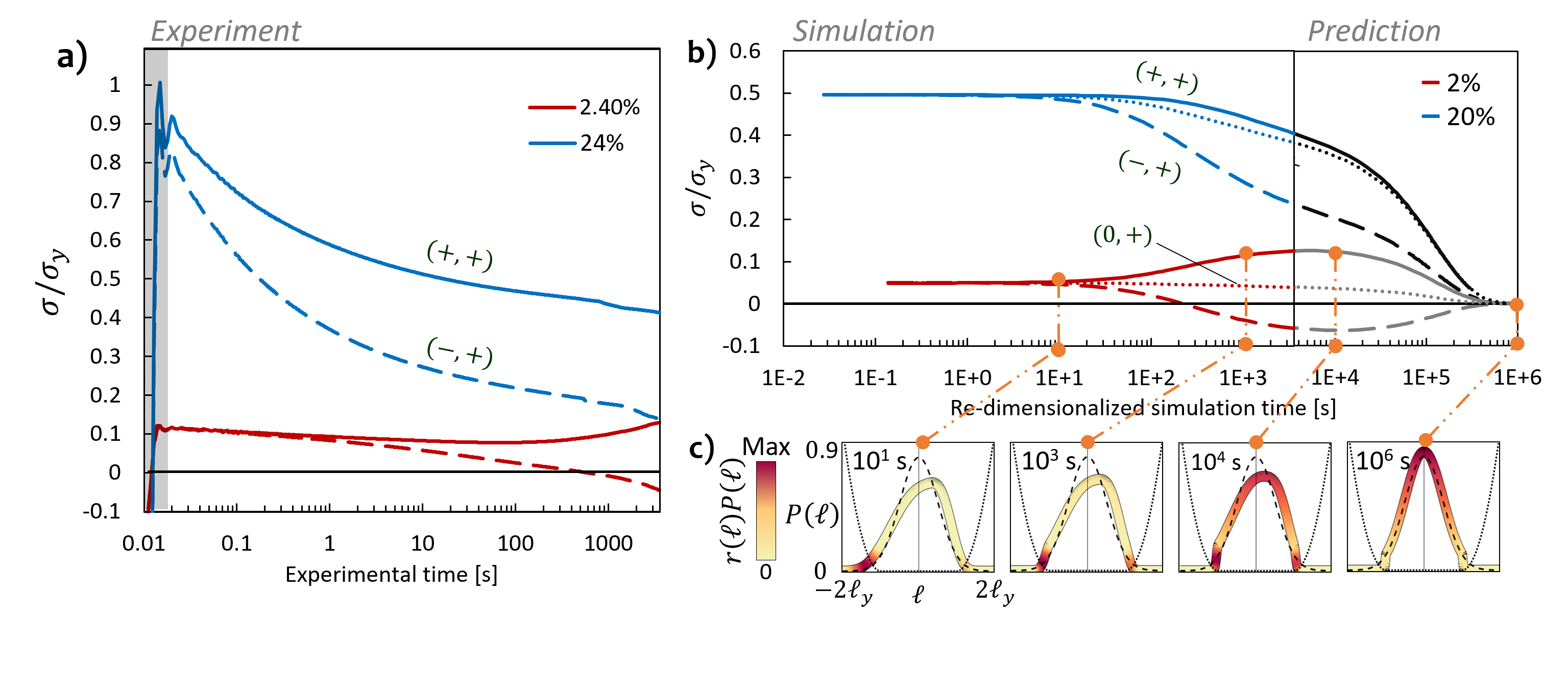}
\caption{\label{fig:SI_directionality_logspace} 
(a) Experimental response of standard Carbopol to stress relaxation after step strains to $2.4\%$ and $24\%$ strain. Steps follow pre-shear in the same direction as the step, $(+,+)$, or opposing direction, $(-,+)$. 
(b) Simulated response of Carbopol-like material in simulation space, including a step without pre-shear, $(0,+)$, for comparison, and showing agreement with observed experiments and terminal relaxation to zero stress. Simulation parameters use a pre-shear rate of $\dot{\gamma}=0.1$, training time $t_{train} = 1$~s, post-hop distribution width $\ell_0/\ell_y=0.5$, yielding fluiditiy $\varepsilon = 0.0034$, and the local yielding rate function $r(\ell)$ uses a power law of $\nu = 2$.
(c) Strain element distributions predicted by the mesoscale statistical model during the relaxation process, color-coded by the yielding rate $r(\ell)P(\ell,t)$ per element $\ell$, showing anisotropy at multiple times. Time was redimensionalized by $\hat{\tau}_{0,Carbopol} = 44$~s.
}
\end{figure*}

We explore the experimental response of standard Carbopol to stress relaxation after step strains to $2.4\%$ and $24\%$ strain in Fig~\ref{fig:SI_directionality_logspace}. Steps follow pre-shear in the same direction as the step, $(+,+)$, or opposing direction, $(-,+)$. 
The simulated response of Carbopol-like material, including a step without pre-shear, $(0, +)$, for comparison, and showing agreement with observed experiments and terminal relaxation to zero stress (Fig~\ref{fig:SI_directionality_logspace}b). 
Strain element distributions predicted by the mesoscale statistical model during the relaxation process, color-coded by the yielding rate $r(\ell)P(\ell,t)$ per element $\ell$, showing anisotropy at multiple times (Fig~\ref{fig:SI_directionality_logspace}c). 

\subsection{Experiment parameters and data handling} \label{SI:expparams}
Typical parameter used are for Step 1: material \textbf{conditioning} at a high constant shear rate ($\hat{\dot{\gamma}} = 10-100$ s$^{-1}$ but the material is generally insensitive to any ``high enough" rates) followed by Step 2: \textbf{training} at a constant zero stress (for $\hat{t}_{train} = 100$~s) and Step 3: a \textbf{reading} period performed as a step strain followed by stress relaxation. 
Within Fig~\ref{fig:protocol}c, experimental data at low stress ($\hat{\sigma}<10$~Pa) and time ($\hat{t}<7$~s) was post-processed by an averaging filter after measurement to smooth, similarly for 
Fig~\ref{fig:directionality} for experimental data at low stress ($\hat{\sigma}<0.5$~Pa) and time ($\hat{t}<7$~s) and for 
Fig~\ref{fig:EM_Carbopol_alone} for experimental data at low stress ($\hat{\sigma}<1$~Pa) and time ($\hat{t}<7$~s).

\subsection{Simulation parameters} \label{SI:modelparams}
\textcolor{black}{The full detail of model parameters and unlisted experimental parameters used within each included figure is listed in Table~\ref{table:parameters}. 
All simulations use a protocol with material \textbf{conditioning} pre-shear $\dot{\gamma} = 0.01$ until steady state, with the exception of experiments for Figs.~\ref{fig:directionality} and \ref{fig:SI_directionality_logspace} which also used a reversed-direction pre-shear $\dot{\gamma} = -0.01$ until steady state, which is marked explicitly in the text for each case by notation $(-, +)$ where the order corresponds to the relative directions of $(\dot{\gamma}, \gamma_{step})$, with the final stress read as positive in the $+$ direction. When not specifically labeled, all terms are $+$. The curves displayed in Fig~\ref{fig:multipart_flow_curve} are an analytical solution and relevant parameters are described in the accompanying text.}

\begin{table}[h!]
\begin{tabular}{ l l l r c r }
 Figure                     & $\ell_0/\ell_y$  & $\varepsilon$     & $\nu$     & $\hat{\tau}_0$ & 
$t_{train}$\\ 
 \hline
 Fig~\ref{fig:protocol}b,d, ~\ref{fig:kymograph}   & 0.8     & 0.007             & 2          & 22 s            & 1 \\
 Fig~\ref{fig:flow_curves}a    & 0.65     & 0             & 0-3          & --            & 1 \\
 Fig~\ref{fig:flow_curves}b    & 0.65     & 0             & 2          & --            & 1 \\
 Fig~\ref{fig:directionality}b    & 0.35     & 0.0005             & 2          & 22 s            & 1 \\
 Fig~\ref{fig:SI_model-vs-experiments}a, b    & 0.65     & 0             & 0-2          & --            & 1 \\
 Fig~\ref{fig:EM_Carbopol_alone}c    & 0.65     & 0.01             & 2         & --            & 1 \\
 Fig~\ref{fig:SI_simulation-sweep}a    & 0.03-2     & 0             & 2          & --            & 1 \\
 Fig~\ref{fig:SI_simulation-sweep}b    & 0.65     & 0             & 2          & --            & 0.1-10 \\
 Fig~\ref{fig:SI_simulation-sweep}c, d    & 0.65     & 0, $10^{-3}$-$10^{-1}$          & 2          & --            & 1 \\
 Fig~\ref{fig:SI_directionality_logspace}    & 0.5     & 0.0034             & 2          & 44 s            & 1 \\
\end{tabular}
\caption{\textcolor{black}{Simulation parameters used in the experiments for each figure including the post-hop distribution width $\ell_0$, background fluidity $\varepsilon$, yielding well exponent $\nu$, simulation time dimension $\hat{\tau}_{0}$, and training time $t_{train}$.}}
\label{table:parameters}
\end{table}

\end{document}